\documentstyle[preprint,aps,eqsecnum]{revtex}

\tighten

\begin{document}

\preprint{\parbox[b]{3.3cm}{NIKHEF-94-P7\\ hep-ph/9408305}}

\draft

\title{Polarized twist-three distributions $g_T$ and $h_L$\\ and the role
of intrinsic transverse momentum}

\author{R.D. Tangerman and P.J. Mulders\thanks{Also at Physics Department,
Free University, NL-1081~HV Amsterdam, The Netherlands.}}

\address{National Institute for Nuclear Physics and High Energy Physics
(NIKHEF-K),\\ P.O. Box 41882, NL-1009~DB Amsterdam, The Netherlands }

\date{August 1994}

\maketitle

\begin{abstract}

In a nonstandard way we split up the polarized quark distributions $g_T$ and
$h_L$ into their twist-two, quark-mass, and interaction-dependent parts,
emphasizing the sensitivity to quark intrinsic transverse momentum.
We show how to derive the Burkhardt-Cottingham sum rule in this
approach and derive a similar sum rule for the chiral-odd distribution $h_2$.
The effect of intrinsic transverse momentum in experimental
observables is illustrated in the calculation of the ${\cal O}(1/Q)$
double-spin asymmetry $A_{LT}$ in Drell-Yan scattering.

\end{abstract}


\section{Introduction}  \label{sec1}

Stimulated by recent~\cite{ESMC} and planned~\cite{RHIC}
experiments with polarized electron and proton beams, spin-dependent
distribution functions have received a lot of attention.
Due to increasing accuracy, measurements of higher-twist
distributions seem feasible in the near future. They are particularly
interesting, since they contain valuable information on quark-gluon
correlations.
The goal of this paper is to analyze the role of quark
transverse momentum in the twist-three distributions\footnote{%
To avoid confusion, we emphasize that we are sometimes using names
reserved for the structure functions measured in inclusive
deep-inelastic lepton-hadron scattering for quark
distributions. In fact, the distributions
should be carrying a quark (or antiquark) flavor index,
which is often suppressed to simplify the formulas.
At the end the structure functions are obtained
as a simple weighted sum over quark distributions.}
$g_T(x)$ and $h_L(x)$, or equivalently the linear combinations
$g_2(x)=g_T(x)-g_1(x)$ and $h_2(x)=2h_L(x)-2h_1(x)$.

At leading-order level, one may consider transverse-momentum-dependent
distributions which can be measured
in Drell-Yan (DY) scattering at small but fixed $Q_T$~\cite{rals79,tang94}.
In deep inelastic lepton-hadron scattering (DIS) or in DY integrated over
$Q_T$ one measures
the distributions integrated over transverse momentum; these are
the twist-two unpolarized momentum distribution $f_1(x)$ and
two twist-two spin-dependent distributions, the helicity distribution $g_1(x)$
and the transverse spin distribution $h_1(x)$. At leading order it does not
matter what the transverse momentum dependence is;
the observables (i.e., structure
functions and asymmetries) do not change if one assumes zero transverse
momentum [i.e., $\delta(\bbox{k}_T^2)$] distributions.
At subleading ${\cal O}(1/Q)$, the situation is less simple.
As was demonstrated in
Ref.~\cite{jaff91a}, for instance, assuming free quarks without transverse
momentum gives the non-physical result $g_2(x)=0$.
Thus, even though $g_T$ and $h_L$ do not depend explicitly on $\bbox{k}_T$,
assuming zero transverse momentum may considerably alter their shape (in $x$).
We will make these statements more precise in due course.

First, let us re\-call some facts about the trans\-verse spin-de\-pen\-dent
dis\-tri\-bu\-tion $g_T(x)$ \cite{jaff91a,jaff90}.
Using operator-product-expansion (OPE) techniques, one shows that $g_T(x)$ can
be decomposed in three pieces, a twist-two part depending on $g_1(x)$, also
known as the Wandzura-Wilczek (WW) term~\cite{wand77}, a quark-mass part
depending on $h_1(x)$~\cite{koda79}, and an explicitly interaction-dependent
part. The latter is the most interesting, since it is sensitive to
off-shellness and confinement effects.
Instead of using the OPE, we rederive the different
terms, using nonlocal matrix elements only. In this approach the transverse
momentum of quarks is an essential ingredient.  The nonlocal matrix element
that fixes the interaction-dependent part contains two good quark fields
and one transverse gluon field, so that it can be given a clear parton
interpretation~\cite{mank91a}. Of course, if moments are taken, one
should recover the OPE local matrix elements.
In this formalism it is also straightforward to derive the Burkhardt-Cottingham
(BC) sum rule $\int_0^1 dx\, g_2(x)=0$~\cite{burk70}, although one must make
assumptions. The status of the BC sum rule is less rigorous than, for
instance, the Bjorken sum rule. It was claimed that in perturbation theory
it is valid at least up to ${\cal O}(\alpha_s)$~\cite{anto81}, although
recently this claim was disputed~\cite{mert93}.

The twist-three distribution function $h_L(x)$~\cite{jaff91b} is less
well-known. It does not occur in the inclusive DIS
cross section other than multiplied with a quark-mass factor
(like $h_1(x)$, it is chirally odd).
Nonetheless, one can analyze it using the OPE-based techniques~\cite{jaff92}.
Again one finds three parts; a twist-two part depending on $h_1(x)$, a
quark-mass part depending on $g_1(x)$, and an interaction-dependent part.
The interaction dependent part can also be given a parton
interpretation\cite{bere94}.
Using the same techniques as for $g_T$, we derive within our approach
these parts, expressing them
in a finite number of nonlocal matrix elements.
We derive a sum rule very similar
to the BC sum rule, namely, $\int_0^1 dx\, h_2(x)=0$.

In order to support our claim that explicit treatment of quark transverse
momenta is important at twist-three, we consider
$Q_T$-averaged polarized DY scattering. Jaffe and Ji derived an
${\cal O}(1/Q)$ asymmetry, $A_{LT}$, for longitudinally-transversely polarized
DY scattering~\cite{jaff92}.
In their calculation they assumed zero quark transverse momentum. Without
this assumption we find modifications in the
asymmetry. To estimate the difference between the two results
we use bag-model distributions. The difference turns out to be considerable.

This paper has the following structure. In Sec.~\ref{sec2}, we analyze the
content of $g_T(x)$ and $h_L(x)$, using Lorentz symmetry, discrete symmetries
and the QCD equations of motion. We decompose the distributions in the three
different parts, and derive the sum rules. In Sec.~\ref{sec3} we
calculate the polarized DIS hadron tensor, emphasising the role
of transverse momentum. In Sec.~\ref{sec4}, the more involved but essentially
similar hadronic tensor for polarized DY scattering is considered, from
which follows the double-spin asymmetry.

\section{Analysis}    \label{sec2}

In this section we decompose the polarized twist-three distributions $g_T(x)$
and $h_L(x)$ into their building blocks of different dynamical origin.
Instead of using conventional OPE techniques, we work directly with the
nonlocal quark-quark and quark-gluon-quark matrix elements. In the calculation
of a specific cross section these matrix elements appear as the
non-calculable parts in the diagrammatic expansion of the amplitude, i.e.,
those parts that connect hadron lines to quark or gluon lines.

\subsection{Correlation functions}

First, we will expand the quark-quark and quark-gluon-quark correlation
functions as needed up to ${\cal O}(1/Q)$, using hermiticity, as well as
Lorentz invariance, parity invariance, and time-reversal invariance.

The correlation function describing a quark of flavor $a$ in a spin-$1/2$
hadron $A$, is defined by~\cite{soja}
\begin{equation}\label{corelgen}
(\Phi_{a/A})_{\alpha\beta}( P S;k)=\int \frac{d^4 x}{(2\pi)^4}\, e^{i k\cdot x}
\langle P S |\overline{\psi}_\beta^{(a)}(0)\psi_\alpha^{(a)}(x)|P S\rangle ,
\end{equation}
diagrammatically represented by the blob in
Fig.~\ref{fig:quarkblob}. Here $P$ is the hadron momentum vector ($P^2=M^2$),
and $S$ is its spin vector ($P\cdot S=0$, and $S^2=-1$).
In lightcone variables (for an arbitrary four-vector
$p= [p^-,p^+,\bbox{p}_{T} ]$, with $p^\pm \equiv (p^0 \pm p^3)/\sqrt{2}$)
they read
\begin{eqnarray}
P&=&\left[\frac{M^2}{2P^+},P^+,{\bf 0}_T\right],\label{Pmom}\\
S&=&\left[-\lambda\frac{M}{2P^+},\lambda \frac{P^+}{M},\bbox{S}_T\right],
\label{Svec}
\end{eqnarray}
where $\lambda$ is the helicity, and $\bbox{S}_T=(S^1,S^2)$ is the transverse
polarization. Note: $\lambda^2+\bbox{S}_T^2=-S^2=1$.
The quark momentum is written as
\begin{equation} \label{kmom}
k =\left[ \frac{k^2+\bbox{k}_{T}^2}{2x P^+}, x P^+, \bbox{k}_{T} \right],
\end{equation}
with the lightcone momentum fraction $x=k^+/P^+$.

An important step in our analysis is the expansion of the correlation function
in  terms of all possible Dirac matrices multiplied by scalar functions
depending on $k\cdot P$ and $k^2$, and vanishing when these variables become
larger than a characteristic hadronic scale $\Lambda^2$.
Using the demands of hermiticity, Lorentz, parity, and time-reversal
invariance, one finds~\cite{rals79,tang94}
\begin{eqnarray}
M^3\, \Phi(P S;k) & = &
a_1\, \bbox{1} + (a_2/M)\, {\not\! P} + (a_3/M)\, {\not\! k}
\nonumber\\
&&+ a_4 \,\gamma_5 {\not\! S}
+ (a_5/M)\, \gamma_5 [{\not\! P},{\not\! S}]
+ (a_6/M)\, \gamma_5 [{\not\! k},{\not\! S}]\nonumber \\
& &
+ (a_7/M^2)\, (k\cdot S) \gamma_5  {\not\! P}
+ (a_8/M^2)\, (k\cdot S) \gamma_5 {\not\! k}
+ (a_9/M^3)\, (k\cdot S)\gamma_5  [{\not\! P},{\not\! k}],
\label{ampli}\end{eqnarray}
where the dimensionless amplitudes $a_i(\sigma,\tau)$ are real functions of the
scalar variables
\begin{eqnarray}
&&\sigma\equiv \frac{2k\cdot P}{M^2}=\frac{2k^- P^+}{M^2}+x,\\
&&\tau  \equiv \frac{k^2}{M^2}=x\frac{2k^- P^+}{M^2}-\frac{\bbox{k}_T^2}{M^2}.
\end{eqnarray}
Since the relevant object in deep inelastic processes is the integral over
$k^-$ of $\Phi(PS;k)$, it is convenient to define the projections
(suppressing flavor and hadron labels)
\begin{eqnarray}
\Phi[\Gamma](x,\bbox{k}_T)&\equiv&\frac{1}{2}\int d k^-\,\text{Tr}
\left[\Gamma\,\Phi(k)\right] \nonumber\\
&=&\frac{1}{2} \int \frac{d x^-}{2\pi}\frac{d^2 \bbox{x}_{T}}{(2\pi)^2}\,
\exp[i(xP^+ x^- -\bbox{k}_{T}\cdot\bbox{x}_{T})]\,\langle P S |
\overline{\psi}(0)\Gamma\psi(0,x^-,\bbox{x}_{T}) | P S\rangle  ,
\label{Phiproje}
\end{eqnarray}
with $\Gamma$ taken from the complete set $\{\bbox{1}$,
$i\gamma_5$, $\gamma^\mu$, $\gamma^\mu\gamma_5$,
$i\sigma^{\mu\nu}\gamma_5\}$. Consider for instance the projection for
$\Gamma=\gamma^+$:
\begin{equation}
\Phi[\gamma^+]=
\frac{1}{M^2}\int d\sigma d\tau\, \delta(\tau-x\sigma+x^2+
\frac{\bbox{k}_T^2}{M^2})\left[a_2(\sigma,\tau)+x a_3(\sigma,\tau)\right],
\end{equation}
where we inserted the expansion of the quark correlation function in terms of
amplitudes, Eq.~(\ref{ampli}). Clearly, it is an ${\cal O}(1)$ function, $f_1$,
depending on $x$ and $\bbox{k}_T$.
Similarly, for $\Gamma=\bbox{1}$ one has
\begin{equation}
\Phi[1]=
\frac{M}{P^+}\frac{1}{M^2}\int d\sigma d\tau\, \delta(\tau-x\sigma+x^2+
\frac{\bbox{k}_T^2}{M^2})\,a_1(\sigma,\tau),
\end{equation}
which is again an ${\cal O}(1)$ function, $e$, of $x$ and $\bbox{k}_T^2$, but
multiplied by a factor $M/P^+$.
In deep inelastic processes, characterized by a
large momentum scale $Q$, this factor will give rise to an $M/Q$ suppression.
The following projections are leading (here and in the rest of the paper, the
indices $i$ and $j$ are transverse, i.e., taking the values $1,2$):
\begin{mathletters}\label{tweetwist}
\begin{eqnarray}
& & \Phi[\gamma^+]= f_1(x,\bbox{k}_{T}^2) ,
\\ & & \Phi[\gamma^+ \gamma_5]=  g_{1L}(x,\bbox{k}_{T}^2) \lambda
+g_{1T}(x,\bbox{k}_{T}^2)\frac{\bbox{k}_T\cdot\bbox{S}_{T}}{M},\label{Phigpgf}
\\ & & \Phi[ i \sigma^{i+}\gamma_5 ]=h_{1T}(x,\bbox{k}_{T}^2) \bbox{S}_{T}^i
+ \left[h_{1L}^\perp(x,\bbox{k}_{T}^2)\lambda +h_{1T}^\perp(x,\bbox{k}_{T}^2)
\frac{\bbox{k}_T\cdot\bbox{S}_{T}}{M}\right]\frac{\bbox{k}_{T}^i}{M},
\label{Phisig}
\end{eqnarray}
\end{mathletters}
where we have defined
\begin{mathletters}\label{ampltwee}
\begin{eqnarray}
f_1(x,\bbox{k}_T^2)&=&\frac{1}{M^2}\int d\sigma d\tau\,
\delta(\tau-x\sigma+x^2+
\frac{\bbox{k}_T^2}{M^2})\left[a_2+x a_3\right],\\
g_{1L}(x,\bbox{k}_T^2)&=&\frac{1}{M^2}\int d\sigma d\tau\,
\delta(\tau-x\sigma+x^2+
\frac{\bbox{k}_T^2}{M^2})\left[a_4+(\case{1}{2}\sigma-x)(a_7+xa_8)\right],\\
g_{1T}(x,\bbox{k}_T^2)&=&\frac{1}{M^2}\int d\sigma d\tau\,
\delta(\tau-x\sigma+x^2+
\frac{\bbox{k}_T^2}{M^2})\left[-a_7-xa_8\right],\label{g1Tampli}\\
h_{1T}(x,\bbox{k}_T^2)&=&\frac{1}{M^2}\int d\sigma d\tau\,
\delta(\tau-x\sigma+x^2+
\frac{\bbox{k}_T^2}{M^2})\left[-2a_5-2xa_6\right],\\
h_{1L}^\perp(x,\bbox{k}_T^2)&=&\frac{1}{M^2}\int d\sigma d\tau\,
\delta(\tau-x\sigma+x^2+
\frac{\bbox{k}_T^2}{M^2})\left[2a_6-2(\case{1}{2}\sigma-x)a_9\right],
\label{h1Lampli}\\
h_{1T}^\perp(x,\bbox{k}_T^2)&=&\frac{1}{M^2}\int d\sigma d\tau\,
\delta(\tau-x\sigma+x^2+ \frac{\bbox{k}_T^2}{M^2})\left[2a_9\right].
\end{eqnarray}
\end{mathletters}
These leading transverse momentum distributions can be directly observed in
DY at measured $Q_T$~\cite{tang94}. For DIS or $Q_T$-averaged DY one
needs only specific $\bbox{k}_T$-integrals of Eqs.~(\ref{tweetwist}).
Particularly, the ones occurring at leading order in $1/Q$ are
\begin{mathletters}\label{goedtwee}
\begin{eqnarray}
&&\int d^2\bbox{k}_T\,\Phi[\gamma^+]=\int d^2\bbox{k}_{T}\,
f_1(x,\bbox{k}_{T}^2)\equiv f_1(x),\\
&&\int d^2\bbox{k}_T\,\Phi[\gamma^+\gamma_5]=
\lambda\int d^2\bbox{k}_{T}\, g_{1L}(x,\bbox{k}_{T}^2)\equiv\lambda\, g_1(x),
\label{g1}\\
&&\int d^2\bbox{k}_T\,\Phi[ i \sigma^{i+}\gamma_5 ]=
\bbox{S}_T^i\int d^2\bbox{k}_{T} \left[h_{1T}(x,\bbox{k}_{T}^2)
+\frac{\bbox{k}_{T}^2}{2M^2}h_{1T}^\perp(x,\bbox{k}_{T}^2)\right]
\equiv\bbox{S}_T^i\, h_1(x). \label{h1}
\end{eqnarray}
\end{mathletters}
Note that $g_{1T}$ and $h^\perp_{1L}$ vanish
because their $\bbox{k}_T$-structure is odd (under the operation
$\bbox{k}_T\rightarrow -\bbox{k}_T$).
We recognize the longitudinal momentum, helicity, and transverse spin
distributions. Expressed in terms of the amplitudes, with the help of
Eqs.~(\ref{ampltwee}), they are
\begin{mathletters}
\begin{eqnarray}
f_1(x)&=&\pi\int
d\sigma d\tau\,\theta(x\sigma - \tau - x^2)\, \left[a_2+x a_3\right],\\
g_1(x)&=&\pi\int
d\sigma d\tau\,\theta(x\sigma - \tau - x^2)\,
\left[a_4+(\case{1}{2}\sigma-x)(a_7+xa_8)\right],\label{g1ampli}\\
h_1(x)&=&\pi\int
d\sigma d\tau\,\theta(x\sigma - \tau - x^2)\,
\left[-2a_5-2xa_6+(x\sigma-\tau-x^2)a_9\right].
\end{eqnarray}
\end{mathletters}
The $\theta$-function together with the spectral condition $(P-k)^2\geq 0$
leads to an integration area as illustrated in Fig.~\ref{fig:physarea}.
Note that for $x=1$ the integration area becomes zero and consequently the
distributions vanish.

For the subleading projections, we will need only the integrals
over $\bbox{k}_T$. Three Dirac projections remain\cite{jaff92},
\begin{mathletters}\label{knar}
\begin{eqnarray}
&&\frac{P^+}{M}\int d^2\bbox{k}_T\,\Phi[\bbox{1}]= e(x),\\
&&\frac{P^+}{M}\int d^2\bbox{k}_T\,
\Phi[\gamma^i\gamma_5]=\bbox{S}_T^i\, g_T(x),
\label{gT}\\
&&\frac{P^+}{M}\int d^2\bbox{k}_T\,
\Phi[i\sigma^{+-}\gamma_5]=\lambda\, h_L(x).
\label{hL}
\end{eqnarray}
\end{mathletters}
These twist-three $x$-dependent distributions in terms of the amplitudes read
\begin{mathletters}
\begin{eqnarray}
e(x)&=&\pi\int
d\sigma d\tau\,\theta(x\sigma - \tau - x^2)\, \left[a_1\right],\\
g_T(x)&=&\pi\int
d\sigma d\tau\,\theta(x\sigma - \tau - x^2)\,
\left[a_4-\case{1}{2}(x\sigma-\tau-x^2)a_8
\right],\label{gTampli}\\
h_L(x)&=&\pi\int
d\sigma d\tau\,\theta(x\sigma - \tau - x^2)\,
\left[-2a_5-\sigma a_6+2(\case{1}{2}\sigma
-x)^2 a_9\right].
\end{eqnarray}
\end{mathletters}
The chirally odd distribution $e(x)$ contributes at
${\cal O}(1/Q^2)$ in unpolarized Drell-Yan scattering~\cite{jaff92}.
The polarized distributions $g_T(x)$ and $h_L(x)$ will
be analyzed further in the next subsections.
In summary, the quark correlation function where the nonlocality is
pure lightlike (and in the $-$-direction) can be parametrized up to and
including twist three as
\begin{eqnarray}
2\int d^2\bbox{k}_T\,dk^-\,\Phi(k)=&&f_1(x)\;\gamma^-
+\lambda g_1(x) \;\gamma_5\gamma^-
+\bbox{S}_T^i h_1(x) \;i\gamma_5\sigma^-\!_i \nonumber\\
&&+\frac{M}{P^+}\left[e(x)\;\bbox{1}
+\bbox{S}_T^i g_T(x) \;\gamma_5\gamma_i
+\lambda h_L(x) \;i\gamma_5\sigma^{+-}\right]\nonumber\\
&&+\text{ higher twist}. \label{Phiexpa}
\end{eqnarray}

At ${\cal O}(1/Q)$, another type of correlation function needs to be
considered, namely, the quark-gluon-quark matrix element
\begin{equation}\label{lijmmat2}
({\cal M}_{a/A})_{\alpha\beta}^i(P S;k,p)=
\int \frac{d^4 x}{(2\pi)^4}\frac{d^4
x'}{(2\pi)^4} \,e^{i k\cdot x}e^{i(p-k)\cdot x'}
\langle P S | \overline{\psi}_\beta^{(a)}(0)g A_T^i (x')
\psi_\alpha^{(a)}(x)|PS\rangle ,
\end{equation}
pictorially represented in Fig.~\ref{fig:gluonblob}.
The demands of hermiticity, parity, and time-reversal invariance lead to the
constraints
\begin{equation}\label{Mcon}
\begin{array}{lcl}
\left[{\cal M}^i(P S;k, p)\right]^\dagger=\gamma^0\,{\cal M}^i(P S;p, k)
\,\gamma^0&\qquad\qquad\quad& [\text{Hermiticity}]\\
\,\,{\cal M}^i(P S;k, p)\,\,= \gamma^0 \,{\cal M}_i(\bar{P}
-\!\!\bar{S};\bar{k},\bar{p}) \,\gamma^0 && [\text{Parity}]\\
\left[{\cal M}^i(P S;k,p)\right]^\ast =\gamma_5 C \,
{\cal M}_i(\bar{P}\bar{S};\bar{k},\bar{p})\, C^\dagger\gamma_5&&
[\text{Time reversal}]
\end{array}
\end{equation}
where $C=i\gamma^2\gamma^0$ is the charge conjugation matrix,
and $\bar k^\mu=k_\mu$.
In deep inelastic processes, at least on the tree level, one is sensitive
to the projections
\begin{eqnarray}
&&{\cal M}^i[\Gamma](x,\bbox{k}_T)\equiv\frac{1}{2}\int d k^- \,d^4p\,
\text{Tr}\left[\Gamma\,{\cal M}^i(k,p)\right]
\nonumber\\
&&\qquad=\frac{1}{2} \int \frac{d x^-}{2\pi}\frac{d^2 \bbox{x}_{T}}{(2\pi)^2}\,
\exp[i(xP^+ x^- -\bbox{k}_{T}\cdot\bbox{x}_{T})]\,\langle P S |
\overline{\psi}(0)g A_T^i(0)\Gamma\psi(0,x^-,\bbox{x}_{T})
| P S\rangle  ,\label{Mproje}
\end{eqnarray}
specifically, at the ${\cal O}(1/Q)$ accuracy level, to the leading ones with
$\Gamma=\gamma^+$, $\gamma^+\gamma_5$, or $i\sigma^{i+}\gamma_5$.
If $Q_T$ is not measured, then at ${\cal O}(1/Q)$ one may also integrate
over $\bbox{k}_T$.
Using the parity and time-reversal constraints of Eq.~(\ref{Mcon}),
we define the following real quark-gluon-quark distributions:
\begin{eqnarray}
\bbox{S}_T^i\,\tilde{g}_T(x)&&\equiv
\frac{1}{2xM}\int d^2\bbox{k}_T\,{\cal M}_j[g_T^{ij}\gamma^+\gamma_5+
i\epsilon_T^{ij}\gamma^+]+\text{h.c.}\nonumber\\
=&&\frac{1}{4xM}\int \frac{d x^-}{2\pi}e^{ixP^+x^-}\,\langle P S |
\overline{\psi}(0)g A_{Tj}(0)
[g_T^{ij}\gamma^+\gamma_5+i\epsilon_T^{ij}\gamma^+]\psi(x^-)
| P S\rangle+\text{h.c.},\label{gTtwiddle}\\
\lambda\, \tilde{h}_{L}(x)&&\equiv
\frac{1}{2xM}\int d^2\bbox{k}_T\,{\cal M}^i[g_{Tij}\,
i\sigma^{j+}\gamma_5]+\text{h.c.}\nonumber\\
=&&\frac{1}{4xM}\int \frac{d x^-}{2\pi}e^{ixP^+x^-}\,\langle P S |
\overline{\psi}(0)g A_{T}^i(0)[g_{Tij}\, i\sigma^{j+}\gamma_5]\psi(x^-)
| P S\rangle+\text{h.c.}
\label{hLtwiddle},
\end{eqnarray}
where $\psi(x^-)$ denotes $\psi(0,x^-,{\bf 0}_T)$ and we have used the tensors
\begin{eqnarray}
&&\label{getrans}
g_T^{\mu\nu}\equiv  g^{\mu\nu} -n_+^{\mu} n_-^{\nu}- n_+^{\nu} n_-^{\mu},\\
&&\label{epstrans}
\epsilon_T^{\mu\nu}\equiv
{\epsilon^{\mu\nu}}_{\rho\sigma}n_{+}^{\rho}n_{-}^{\sigma}.
\end{eqnarray}
The lightlike vector $n_+$  and $n_-$ have only
a non-zero $+$ and $-$ component, respectively,
such that $n_+\cdot n_-=1$. The reason for the suggestive nomenclature
of $\tilde{g}_T(x)$ and $\tilde{h}_L(x)$ will become clear later.

\subsection{Decomposition of $g_T(x)$}

The first important step in the splitting up of $g_T(x)$ into its constituent
parts is the implementation of the QCD equations of motion for the quark
fields $i{\not\! D}\psi=m\psi$~\cite{jaso}.
In lightcone language~\cite{kogu70} this implies the elimination
of the `bad' field $\psi_-\equiv\Lambda_-\psi=(\gamma^+\gamma^-/2)\psi$
occurring
in $g_T(x)$ [Eq.~(\ref{gT})]
in favor of the `good' fields $\psi_+\equiv\Lambda_+\psi
= (\gamma^-\gamma^+/2)\psi$ and $A_T$. Explicitly one has
\begin{eqnarray}
0&=&\int \frac{d^4 x}{(2\pi)^4}e^{ik\cdot x}\langle PS|\overline{\psi}(0)
\Bigl[i{\not\! D}(0)-m\Bigr]i\sigma^{i+}\gamma_5\psi(x)|PS\rangle\\
&=&\int \frac{d^4 x}{(2\pi)^4}e^{ik\cdot x}\langle PS|\overline{\psi}(0)
\Bigl[{\not\! k}+g{\not\! A}(0)-m\Bigr]i\sigma^{i+}\gamma_5\psi(x)|PS\rangle\\
&=&\int \frac{d^4 x}{(2\pi)^4}e^{ik\cdot x}\langle PS|\overline{\psi}(0)
\left[k^+\gamma^-+\Bigl(\bbox{k}_T^j+gA_T^j(0)\Bigr)\gamma_j-m\right]
i\sigma^{i+}\gamma_5\psi(x)|PS\rangle ,
\end{eqnarray}
where translational invariance of the hadron states is used and the lightcone
gauge $A^+=0$ is chosen.
Performing the integral $\case{1}{2}\int dk^-$, one obtains
\begin{eqnarray}
&&k^+\Phi[\gamma^i\gamma_5]+ik^+\epsilon_T^{ij}\Phi[\gamma_j]\nonumber\\
&&\qquad\qquad=k_T^i\Phi[\gamma^+\gamma_5]
+i\epsilon_T^{ij}k_{Tj}\Phi[\gamma^+]
+m\Phi[i\sigma^{i+}\gamma_5]
+{\cal M}_j[g_T^{ij}\gamma^+\gamma_5+i\epsilon_T^{ij}\gamma^+],
\end{eqnarray}
using the definitions~(\ref{Phiproje}) and~(\ref{Mproje}).
Finally, upon taking the real part\footnote{Note that
$\Phi[\Gamma]$ is real if $\Gamma$ is taken from the set $\{\bbox{1}$,
$i\gamma_5$, $\gamma^\mu$, $\gamma^\mu\gamma_5$,
$i\sigma^{\mu\nu}\gamma_5\}$.}, integrating over $\bbox{k}_T$, and using
the parametrizations for the projections given in Eqs.~(\ref{Phigpgf}),%
{}~(\ref{h1}),~(\ref{gT}), and~(\ref{gTtwiddle}), one gets
\begin{equation}
g_T(x)=\int d^2\bbox{k}_T \frac{\bbox{k}_T^2}{2M^2}\frac{g_{1T}
(x,\bbox{k}_T^2)}{x} +\frac{m}{M}
\frac{h_1(x)}{x}+\tilde{g}_T(x).\label{eqgt}
\end{equation}
This result will be used in the next sections, when we will calculate DIS and
DY structure functions.
We make a few remarks. First, at ${\cal O}(1/Q)$ the transverse
covariant derivative inevitably introduces a sensitivity to the
transverse-momentum-dependence comprised in the first term
on the lefthandside of Eq.~(\ref{eqgt}), {\em even\/} if
transverse momentum is not observed directly. Secondly, this term
contains the functions $g_{1T}$, to which one is not
sensitive at leading order if transverse momentum is not observed [see
Eq.~(\ref{g1})]. If this distribution is proportional to a delta
function $\delta(\bbox{k}_T^2)$, the integral vanishes. So if one assumes zero
transverse
momentum, the difference between $g_T(x)$ and $\tilde{g}_T(x)$ is of order
$m/M$. In general, however, this cannot be assumed.

It is possible to re-express the above relation in terms of the
$\bbox{k}_T$-integrated distributions only.
For this one needs to use the expressions in terms of the
amplitudes $a_i(\sigma,\tau)$ given in the preceding subsection
and the following relation for a linear combination of amplitudes
${\cal F}(x,\sigma,\tau)$:
\begin{eqnarray}
&&\int \frac{d^2\bbox{k}_T}{M^2}\frac{\bbox{k}_T^2}{M^2}
\int d\sigma d\tau\, \delta(\tau-x\sigma+x^2+
\frac{\bbox{k}_T^2}{M^2}){\cal F}(x,\sigma,\tau)=\nonumber\\
&&-\pi\int_x^1 dy\int d\sigma d\tau\,
\theta(y\sigma-\tau-y^2)\left[(\sigma-2y){\cal F}(y,\sigma,\tau)
+(y\sigma-\tau-y^2)\frac{\partial{\cal F}}{\partial y}(y,\sigma,\tau)\right],
\label{aap}\end{eqnarray}
which is most easily proven by differentiating both sides with respect to $x$
and using the fact that the integration area vanishes at $x = 1$.
Applying it to the term containing $g_{1T}(x,\bbox{k}_T^2)$ in
Eq.~(\ref{eqgt}), using its expansion in amplitudes
Eq.~(\ref{g1Tampli}), this gives after some algebra
\begin{equation}\label{wwrel}
g_T(x)=\int_x^1
dy\,\frac{g_1(y)}{y} +\frac{m}{M}\left[ \frac{h_1(x)}{x}-\int_x^1
dy\,\frac{h_1(y)}{y^2}\right] +\tilde{g}_T(x)-\int_x^1
dy\,\frac{\tilde{g}_T(y)}{y}.
\end{equation}
The first term was derived by Wandzura and Wilczek
in~\cite{wand77}. The quark-mass terms are chirally even, since
both $m$ and $h_1(x)$ are chirally odd. The last two explicitly
interaction-dependent terms only need the nonlocal matrix element
$\tilde{g}_T(x)$, defined in Eq.~(\ref{gTtwiddle}).

With the relation~(\ref{wwrel}), one easily checks that
\begin{equation}\label{BC}
\int_0^1 dx\, g_2(x)=
\int_0^1 dx\,\left[ g_T(x)-g_1(x)\right]=
\int_0^1 dx\,G(x)-\int_0^1 dx\int_x^1 \frac{dy}{y}G(y),
\end{equation}
with $G(x)\equiv -g_1(x)+(m/M)[h_1(x)/x]+\tilde{g}_T(x)$. Provided
that in Eq.~(\ref{BC}) one may interchange the order of integrations,
this amounts to the Burkhardt-Cottingham sum rule
$\int_0^1 dx\, g_2(x)=0$~\cite{burk70}.
For more details concerning its validity, see
Refs.~\cite{jaff90,jaff91a,mank91b}.

In Ref.~\cite{jaff91a} Jaffe and Ji calculated the contributions of different
dynamical origin in $g_T(x)$ for a massless quark in a nucleon bag.
In Fig.~\ref{fig:gete} their results are plotted. Also plotted is
$\tilde{g}_T(x)$ [the authors call this function $U(x)$].
Note that, as we argued before, the role
of transverse momentum is fairly important, because else $g_T(x)$ would
equal $\tilde{g}_T(x)$, as follows from Eq.~(\ref{eqgt}) with a $\delta
(\bbox{k}_T^2)$ dependence and $m=0$.
Nonzero transverse momentum is an important feature for a confined quark.
Note finally that the BC sum rule in the bag reads
$\int_{-\infty}^\infty dx\, g_2(x)=0$, because violation of translational
invariance leads to a support between $-\infty$ and $\infty$.
See also Fig.~\ref{fig:ghtwee}.

\subsection{Decomposition of $h_L(x)$}

The twist-three distribution $h_L(x)$~\cite{jaff91b} can be
treated in a similar fashion. This time, implementing the QCD equations of
motion, one must consider the matrix element
\begin{equation}
\langle PS|\overline{\psi}(0)\Bigl[i{\not\! D}(0)-m\Bigr]
\gamma^+\gamma_5\psi(x)|PS\rangle =0.
\end{equation}
This leads to the relation
\begin{equation}
h_L(x)=-\int d^2\bbox{k}_T \frac{\bbox{k}_T^2}{M^2}\frac{h^\perp_{1L}
(x,\bbox{k}_T^2)}{x}+\frac{m}{M}\frac{g_1(x)}{x}+\tilde{h}_L(x),
\label{eqhl}
\end{equation}
involving the transverse momentum distribution $h^\perp_{1L}(x,\bbox{k}_T^2)$
to which one is not sensitive at leading order [Eq.~(\ref{h1})].
The $\bbox{k}_T$-integral can be re-expressed as an $x$-integral
by means of the relation~(\ref{aap}) using the amplitude expansion
Eq.~(\ref{h1Lampli}). We obtain for the decomposition of $h_L(x)$,
\begin{equation}\label{jaffrel}
h_L(x)=2x\int_x^1 dy\,\frac{h_1(y)}{y^2} +\frac{m}{M}\left[
\frac{g_1(x)}{x}-2x\int_x^1 dy\,\frac{g_1(y)}{y^3}\right]
+\tilde{h}_L(x)-2x\int_x^1 dy\,\frac{\tilde{h}_L(y)}{y^2}.
\end{equation}
The compact expression of the interaction-dependent part in terms of
$\tilde{h}_L(x)$, defined in Eq.~(\ref{hLtwiddle}), may be a practical one
for investigating models.
Relation~(\ref{jaffrel}) was also derived by
Jaffe and Ji\footnote{We find a discrepancy of a factor of $2$ in the
third terms on the righthandsides of Eq.~(\ref{jaffrel})
and Eq.~(56) of Ref.~\cite{jaff92}.
The lower integration limit and the power in the integrand in the latter case
are obviously misprints, Cf. Eq.~(51) of the same paper.
During their calculation, Jaffe and Ji also introduce a function
$\tilde{h}_L(x)$, which is not the same as ours.
Because in their case it is an auxiliary function,
which doesn't occur anymore later, and as good names are scarce, we
considered it safe to re-use the name.}
using OPE techniques~\cite{jaff92}.

With the above relation in hand, we consider
\begin{equation}
\int_0^1 dx\, h_2(x)=
2\int_0^1 dx\,\left[ h_L(x)-h_1(x)\right]=
\int_0^1 dx\, H(x)-\int_0^1 dx\, 2x\int_x^1 \frac{dy}{y^2}H(y),
\end{equation}
with $H(x)/2\equiv -h_1(x)+(m/M)[g_1(x)/x]+\tilde{h}_L(x)$. Provided
that the order of $x$- and $y$-integration of the second term may be
interchanged, we find the sum rule $\int_0^1 dx\, h_2(x)=0$. The validity of
this sum rule, of which no reference is known to us, crucially depends on the
small-$x$ behavior of the function $H(x)$.

In Ref.~\cite{jaff92} the different contributions to $h_L(x)$ were calculated
for a massless quark in the bag. Their results are plotted in
Fig.~\ref{fig:hael}.
A comparison with the bag-model $\tilde{h}_L(x)$ in the same figure
shows that the zero-quark-mass zero-transverse-momentum relation
$h_L(x)=\tilde{h}_L(x)$, following from Eq.~(\ref{eqhl}), is severely
violated. Remarkably, $h_2(x)=2g_2(x)$ in the bag (see Fig.~\ref{fig:ghtwee}),
and it satisfies the above sum rule, provided that one extends the integration
region to the full line $(-\infty,\infty)$.

\section{Polarized DIS} \label{sec3}

In the following two sections we address the question how the different
distributions of the preceding section occur in the physical observables.
In this section the polarized deep inelastic cross section
through order $1/Q$~\cite{efra} is calculated as an illustration of the more
complicated DY calculation. The diagrammatical method by Ellis, Furma\'nski,
and Petronzio~\cite{ellr} is used.

First, let us discuss the kinematics. The target hadron has momentum and spin
vectors,
given in lightcone representation in Eqs.~(\ref{Pmom}) and~(\ref{Svec}).
The incoming virtual photon is assumed to have momentum $q$ such that
$-q^2\equiv Q^2$ is large, but $x_{bj}\equiv Q^2/2P\cdot q$ is constant,
and can be chosen\footnote{Here and in the
following we will denote by `$\approx$' that we have neglected
${\cal O}(1/Q^2)$ contributions.}
\begin{equation}\label{qmom}
q\approx \left[ \frac{Q^2}{2x_{bj} P^+},-x_{bj} P^+,{\bf 0}_T\right].
\end{equation}
Instead of working with explicit vectors $P$ and $q$, it is often more
convenient to express them in terms of the
lightlike vectors $n_+$ and $n_-$, defined as
\begin{mathletters}\label{defnpm}
\begin{eqnarray}
&&n_+=\left[0,\kappa,{\bf 0}_T\right]\approx\frac{\sqrt{2}}{Q}x_{bj} P,\\
&&n_-=\left[\frac{1}{\kappa},0,{\bf 0}_T\right]
\approx\frac{\sqrt{2}}{Q}(q+x_{bj} P),
\end{eqnarray}
\end{mathletters}
where the parameter $\kappa=\sqrt{2}\, x_{bj} P^+/Q$.

Consider the quark Born diagram of Fig.~\ref{fig:DISBorn} (the antiquark
diagrams can be obtained by crossing). It stands for
\begin{equation}\label{DISBorn}
2M\,W^{\mu\nu}_B
=\sum_a e_a^2\int d^4 k\, \delta\left((k+q)^2-m^2\right)
\text{Tr}\left[\Phi_a(k)\,\gamma^\mu\,
({\not\! k}+{\not\! q}+m)\,\gamma^\nu\right].
\end{equation}
Where the sum runs over quark flavors $a$ with charge $e_a$ in units of $e$.
Using the lightcone representations for $k$ and $q$, it is easily checked that
\begin{equation}\label{delap}
\delta\left((k+q)^2-m^2\right)\approx \frac{\delta(k^++q^+)}{2q^-},
\end{equation}
equating $x$ to $x_{bj}$.
Furthermore one can write
\begin{equation}\label{polsum}
{\not\! k}+{\not\! q}+m\approx q^-\gamma^++\bbox{k}_T^i\gamma_i+m.
\end{equation}
Note that the only $k^-$-dependence of the integrand in Eq.~(\ref{DISBorn})
resides in $\Phi(k)$, hence indeed we are sensitive to $\int dk^-\,\Phi(k)$.
 From explicit calculation (or from dimensional arguments) the ${\cal O}(Q^0)$
result is obtained from the first term in Eq.~(\ref{polsum}) resulting in
an expression that only contains the leading projections $\Phi[\Gamma]$
of Eqs.~(\ref{tweetwist}) integrated over $\bbox{k}_T$.
Therefore, at leading order, one is sensitive to the (chirally even)
distributions $f_1(x)$ and $g_1(x)$, defined in Eqs.~(\ref{goedtwee}).
In fact,
\begin{equation}
2M\,W^{\mu\nu}_{\text{B}0}=
\sum_a e_a^2 \left[-g_T^{\mu\nu}\;f_1^a(x_{bj})+i\epsilon_T^{\mu\nu}\;
g_1^a(x_{bj})\right],  \label{DISB0}
\end{equation}
where the transverse tensors are defined in Eqs.~(\ref{getrans})
and~(\ref{epstrans}). This leading order result is electromagnetically
current conserving, that is $q_\mu W^{\mu\nu}_{\text{B}0}=0$.

Turning to the ${\cal O}(1/Q)$ contributions, we find that they
either come from the first term in Eq.~(\ref{polsum})
combined with the subleading projections,
or from the combination of the last two terms in Eq.~(\ref{polsum}) and
the leading projections. For the former combinations the same argument holds
concerning the $\bbox{k}_T$-integral, hence we only need the $x$-dependent
distributions of Eqs.~(\ref{knar}). For the latter combinations the argument
goes for the $m$-terms, but the second term of Eq.~(\ref{polsum}) selects the
transverse momentum distributions that multiply a $\bbox{k}_T$-{\em odd\/}
structure, i.e., $g_{1T}(x,\bbox{k}_T^2)$ and $h^\perp_{1L}(x,\bbox{k}_T^2)$.
Thus, after performing the trace and the integrations, we end up with the
${\cal O}(1/Q)$-part of the Born diagram
\begin{eqnarray}
2M\,W^{\mu\nu}_{\text{B}1}=\frac{M}{Q}\sum_a e_a^2 \biggl\{ &&
\sqrt{2}i{\epsilon^{\mu\nu}}_{\rho\sigma}S_T^\rho n_-^\sigma\;x_{bj}
g_T^a(x_{bj})\nonumber\\
&&-\sqrt{2}i{\epsilon^{\mu\nu}}_{\rho\sigma}S_T^\rho n_+^\sigma\biggl[
\int d^2\bbox{k}_T\frac{\bbox{k}_T^2}{2M^2}g_{1T}^a(x_{bj},\bbox{k}_T^2)
+\frac{m}{M}h_1^a(x_{bj})\biggr]\biggr\},
\end{eqnarray}
where $S_T^\mu\equiv g_T^{\mu\nu}S_\nu$.
We obtain a more physical picture if we eliminate $g_T(x)$ by means of the
relation~(\ref{eqgt}) found in the preceding section, so that
the result contains only distributions which are matrix
elements of the independent `good' fields $\psi_+$ and $A_T$.
We get
\begin{eqnarray}
2M\,W^{\mu\nu}_{\text{B}1}=\frac{M}{Q}\sum_a e_a^2 \biggl\{
&&2i{\epsilon^{\mu\nu}}_{\rho\sigma}S_T^\rho \hat{q}^\sigma\biggl[
\int d^2\bbox{k}_T\frac{\bbox{k}_T^2}{2M^2}g_{1T}^a(x_{bj},\bbox{k}_T^2)
+\frac{m}{M}h_1^a(x_{bj})\biggr]
\nonumber\\&&
+\sqrt{2}i{\epsilon^{\mu\nu}}_{\rho\sigma}S_T^\rho n_-^\sigma\;x_{bj}
\tilde{g}_T^a(x_{bj})\biggr\},  \label{B1}
\end{eqnarray}
where $\hat{q}\equiv q/Q\approx (n_- - n_+)/\sqrt{2}$.
Note that the first two terms are current conserving in the sense that they
give $0$ if contracted with $q$. This is as expected, since for the
theory without interactions, i.e., without gluons, one has $\tilde{g}_T(x)=0$,
and the Born diagram is the complete result. In QCD, however, one must
include the gluon diagrams of Fig.~\ref{fig:DISgluons} which contribute at
${\cal O}(1/Q)$ as well.
Written out in full, they are
\begin{eqnarray}
2M\,W_{\text{gluon}}^{\mu\nu} = \sum_a e_a^2 &&
\int d^4k\, d^4 p\,\delta\left((k+q)^2-m^2\right)\nonumber\\
&&\times\Biggl\{
\text{Tr}\left[{\cal M}^i_a (k,p)\, \gamma^\mu\, \frac{i}{{\not\! p}
+{\not\! q}-m}\, i\gamma_i\, ({\not\! k}+{\not\! q}+m)\,\gamma^\nu\right]
\nonumber\\
&&\quad +\text{Tr}\left[{\cal M}^i_a(p,k)\, \gamma^\mu\,  ({\not\! k}
+{\not\! q}+m)\, i\gamma_i\, \frac{i}{{\not\! p}
+{\not\! q}-m}\, \gamma^\nu\right] \Biggr\}.
\end{eqnarray}
The delta function can be approximated as in Eq.~(\ref{delap}).
The extra parton momentum again has $p^2,\bbox{p}_T^2 \ll Q^2$. At
${\cal O}(1/Q)$ accuracy, the polarization sum
may be replaced by $q^- \gamma^+$.
Realizing that $\gamma^+\gamma_i\gamma^+ = 0$, one can effectively use for
the hard propagator
\begin{equation}\label{DISprop}
\frac{1}{{\not\! p}+{\not\! q}-m}\rightarrow
\frac{\gamma^-}{2q^-} = \frac{{\not\! n_+}}{Q\sqrt{2}}.
\end{equation}
Since this does not depend on $p$ anymore, the corresponding integral works on
the ${\cal M}$ directly and we are sensitive
to the projections~(\ref{Mproje}) or the complex conjugates of them
(using the hermiticity condition~(\ref{Mcon}) to exchange
the order of the arguments).
Particularly, in leading order, one has only the projections involving
$\gamma^+$, $\gamma^+\gamma_5$, or $i\sigma^{i+}\gamma_5$.
Also, in leading order, one only needs the $\bbox{k}_T$-integrated projections.
Finally, it turns out that exactly the combination~(\ref{gTtwiddle}),
parametrized by $\tilde{g}_T(x)$, occurs;
\begin{equation}
2M\,W^{\mu\nu}_{\text{gluon}}=-\frac{M}{Q}\sum_a e_a^2
\;\sqrt{2}i{\epsilon^{\mu\nu}}_{\rho\sigma}S_T^\rho n_+^\sigma\;x_{bj}
\tilde{g}_T^a(x_{bj})+{\cal O}\left(\frac{1}{Q^2}\right).
\end{equation}
As expected, this combines with the Born diagram~(\ref{B1}), such that the
total ${\cal O}(1/Q)$-result
\begin{eqnarray}
2M (W^{\mu\nu}_{\text{B}1}+W^{\mu\nu}_{\text{gluon}})&\approx&\frac{2M}{Q}
i{\epsilon^{\mu\nu}}_{\rho\sigma}S_T^\rho \hat{q}^\sigma
\nonumber\\&&\times\sum_a e_a^2\biggl[
\int d^2\bbox{k}_T\frac{\bbox{k}_T^2}{2M^2}g_{1T}^a(x_{bj},\bbox{k}_T^2)
+\frac{m}{M}h_1^a(x_{bj})+x_{bj}\tilde{g}_T^a(x_{bj})\biggr]\\
&=&\frac{2x_{bj} M}{Q}
i{\epsilon^{\mu\nu}}_{\rho\sigma}S_T^\rho \hat{q}^\sigma
\;\sum_a e_a^2\;  g_T^a(x_{bj}),  \label{DISB1}
\end{eqnarray}
is current conserving.
Observe that the final result can be expressed solely
in terms of the distributions $g_T^a(x)=g_1^a(x)+g_2^a(x)$
It will turn out that the DY-case is not that simple.
The flavor sums in the zeroth order result Eq.~(\ref{DISB0}) and
the first order result Eq.~(\ref{DISB1}) may be extended to
include antiquarks, provided that one replaces
$f^a\rightarrow f^{\bar{a}}$, and $g^a\rightarrow -g^{\bar{a}}$,
in accordance with charge conjugation invariance.
Using the standard decomposition in structure functions one sees that
$F_1(x_{bj})=(1/2)\sum_a e_a^2
\left[ f_1^a(x_{bj})+f_1^{\bar{a}}(x_{bj})\right]$, and
$g_i(x_{bj})=(1/2)\sum_a e_a^2
\left[ g_i^a(x_{bj})-g_i^{\bar{a}}(x_{bj})\right]$ with $i=1,2$.

\section{Polarized DY} \label{sec4}

\subsection{Kinematics}

The DY hadron tensor is a bit more involved, since it contains two hadronic
blobs. The principle steps, however, resemble closely those of the preceding
section.
First, some kinematical preliminary remarks.
In the Drell-Yan process or massive-dilepton production, $A+B\rightarrow \ell
+\bar{\ell}+X$, the two spin-$\frac{1}{2}$ hadrons have
momenta $P_A$ and $P_B$, respectively, and are assumed to be pure spin states.
That is, their spin vectors satisfy $S_A^2=S_B^2=-1$. The measured lepton pair
with momenta $k_1$ and $k_2$ has total momentum $q=k_1+k_2$.
The relevant lightcone-coordinate representations are listed below,
\begin{eqnarray}
 P_A &=& \left[ \frac{M_A^2}{2P_A^+},  P_A^+ , \bbox{0}_{T} \right],
\label{pamom}\\
 P_B &=& \left[ P_B^-, \frac{M_B^2}{2P_B^-},\bbox{0}_{T} \right],\label{pbmom
}\\
 S_A &=& \left[ -\lambda_A\frac{M_A}{2P_A^+},\lambda_A \frac{ P_A^+ }{M_A},
\bbox{S}_{AT}\right],
\label{samom}\\
 S_B &=& \left[\lambda_B \frac{P_B^-}{M_B}  ,-\lambda_B \frac{ M_B}{2P_B^-} ,
\bbox{S}_{BT}\right],
\label{sbmom}\\
 q &=& \left[ x_B P_B^-,  x_A P_A^+ , \bbox{q}_{T} \right].
\label{qmomentum}
\end{eqnarray}
We work in the Drell-Yan limit where $q^2\equiv Q^2$ and $s=(P_A+P_B)^2$
become large with fixed ratio $\tau=Q^2/s$. Also we consider only dileptons
with transverse momentum $Q_T^2\equiv \bbox{q}_T^2\lesssim\Lambda^2$.
We again define lightlike vectors $n_+$ and $n_-$ as in Eqs.~(\ref{defnpm}),
with $\kappa =\sqrt{2}\, x_AP_A^+/Q\approx Q/(\sqrt{2}\, x_B P_B^-)$.
The transverse tensors $g_T^{\mu\nu}$ and $\epsilon_T^{\mu\nu}$ are
defined in terms of them as before [Eqs.~(\ref{getrans}) and~(\ref{epstrans})].
Important is to note the precise meaning of {\em transverse\/} here.
A transverse vector $a_T^\mu\equiv g_T^{\mu\nu}a_\nu$ has both $a_T\cdot P_A=0$
and $a_T\cdot P_B=0$. In general the photon {\em does\/} have
transverse components.

The dominant elementary process underlying the reaction
is the annihilation of a quark (antiquark) of hadron~$A$ by an antiquark
(quark) of hadron~$B$ into a massive photon of momentum $q$ which
subsequently decays into the lepton pair.
The cross section can be written as
\begin{equation}\label{cross}
\frac{d\sigma}{d^4 q d\Omega}=\frac{\alpha^2}{2s\,Q^4} L_{\mu\nu}W^{\mu\nu},
\end{equation}
where the hadron tensor
\begin{equation}
W^{\mu\nu}( q; P_A S_A; P_B S_B )=
\int \frac{d^4 x}{(2\pi)^4}\; e^{i q\cdot x} \langle  P_A S_A; P_B S_B |\,
[J^\mu (0), J^\nu (x)]\, | P_A S_A; P_B S_B \rangle .  \label{hadrten}
\end{equation}
is contracted with the lepton tensor
\begin{equation}\label{leptten}
L^{\mu\nu} =
2\,k_1^{\mu}k_2^{\nu}+ 2\,k_2^{\mu}k_1^{\nu}- Q^2\,g^{\mu\nu}  .
\end{equation}
Since the latter is symmetric, we will henceforth discard the antisymmetric
part of the DY hadron tensor. One can rewrite the lepton tensor as
\(L^{\mu\nu}=-Q^2(g^{\mu\nu}-\hat{q}^\mu\hat{q}^\nu+\hat{l}^\mu\hat{l}^\nu )\),
where $\hat{l}\equiv (k_1-k_2)/Q$, satisfying $\hat{l}\cdot\hat{q}=0$
and $\hat{l}^2=-1$, defines the lepton axis.
The angles $\theta$ and $\phi$ are those of the lepton axis measured in a
particular dimuon rest frame ${\cal O}'$ defined by Collins and
Soper~\cite{coll77}. The polar angle is defined with respect to
\begin{equation}
Z^\mu  \equiv  \frac{P_B\cdot q}{P_B\cdot P_A}\, P^\mu_A
              -\frac{P_A\cdot q}{P_A\cdot P_B}\, P^\mu_B , \label{colsop}
\end{equation}
or rather its normalized version $\hat{z}^\mu=Z^\mu/\sqrt{-Z^2}$.
Note that $q\cdot Z=0$ and that $Z$ does not have transverse components.
The azimuthal angle is fixed with respect to $q_\perp\equiv
q_T-(q_T\cdot\hat{q})\hat{q}$, with $\hat{q}^\mu=q^\mu/Q$.
The four-vector $q_\perp$ is an example of what we call a
{\em perpendicular\/} vector~\cite{tang94}. Defining
\begin{equation}\label{geperp}
g_\perp^{\mu\nu}\equiv g^{\mu\nu}-\hat{q}^{\mu}\hat{q}^{\nu}
+\hat{z}^{\mu}\hat{z}^{\nu},
\end{equation}
a perpendicular four-vector $a_\perp^\mu\equiv
g_\perp^{\mu\nu}a_{T\nu}$ satisfies
$a_\perp\cdot \hat{q}=a_\perp\cdot\hat{z}=0$.
For completeness we introduce
\begin{equation}
\epsilon_\perp^{\mu\nu}\equiv {\epsilon^{\mu\nu}}_{\rho\sigma}
\hat{z}^{\rho}\hat{q}^{\sigma}.
\end{equation}
The difference between `transverse' and `perpendicular' is of ${\cal O}(1/Q)$,
\begin{eqnarray}
&& g_T^{\mu\nu}\approx g_\perp^{\mu\nu}+
\frac{(\hat{q}^{\mu} q_\perp^{\nu}+ \hat{q}^{\nu} q_\perp^{\mu}) }{Q},
\label{gtgp}\\
&& \epsilon_T^{\mu\nu}\approx\epsilon_\perp^{\mu\nu}
-\frac{{\epsilon^{\mu\nu}}_{\rho\sigma}\hat{z}^{\rho}q_\perp^{\sigma}}{Q},\\
&& n_+ \approx \frac{1}{\sqrt{2}}
\left(\hat{q} + \hat{z} -\frac{q_\perp}{Q} \right),\\
&& n_- \approx \frac{1}{\sqrt{2}}
\left(\hat{q} - \hat{z} -\frac{q_\perp}{Q}\right),\\
\label{Pvector}
&& a_T \approx a_\perp + \frac{a_\perp\cdot q_\perp}{Q}\,\hat{q},\\
\label{Pinprod}
&& a_T\cdot b_T\approx a_\perp\cdot b_\perp,
\end{eqnarray}
provided that $a_T\cdot q_T$, $b_T\cdot q_T={\cal O}(1)$.

\subsection{Hadron tensor}

We first calculate the DY quark Born diagram of Fig.~\ref{fig:DYBorn}.
It gives
\begin{equation}\label{Born}
W_{\text{B}}^{\mu\nu} =\frac{1}{3}\sum_{a,b}\delta_{b\bar{a}} e^2_a
\int d^4k_a\, d^4k_b\;\delta^4(k_a + k_b -q)
\;\text{Tr} \left[ \Phi_a (k_a)\; \gamma^\mu \;
\Phi_b(k_b)\; \gamma^\nu \right] ,
\end{equation}
where the antiquark correlation function $\Phi_{\bar{a}}$ is defined as
in~(\ref{corelgen}), but with
$\psi^{(a)}\leftrightarrow\overline{\psi}^{(a)}$.
The factor $1/3$ is a color factor.
The quark and antiquark momentum vectors read in lightcone coordinates
\begin{eqnarray}
k_a &=&\left[ \frac{k_a^2+\bbox{k}_{aT}^2}{2x_a P_A^+}, x_a P_A^+,
\bbox{k}_{aT} \right],\\
k_b &=&\left[x_b P_B^-, \frac{k_b^2+\bbox{k}_{bT}^2}{2x_b P_B^-},
\bbox{k}_{bT} \right],
\end{eqnarray}
and we assume that $k_a^2$, $\bbox{k}_{aT}^2$, $k_b^2$, and
$\bbox{k}_{bT}^2$ all are of ${\cal O}(1)$. Since then $k_b^+\ll k_a^+$ and
$k_a^-\ll k_b^-$, one can approximate the delta function,
\begin{equation}\label{delta}
\delta^4(k_a + k_b -q)\approx\delta(k_a^+-q^+)\delta(k_b^- -q^-)\delta^2
(\bbox{k}_{aT}+\bbox{k}_{bT}-\bbox{q}_{T}),
\end{equation}
and Eq.~(\ref{Born}) reduces to
\begin{eqnarray}
W_{\text{B}}^{\mu\nu} \approx\frac{1}{3}\sum_{a,b}\delta_{b\bar{a}} e^2_a &&
\int d^2\bbox{k}_{aT}\, d^2\bbox{k}_{bT}\;\delta^2(\bbox{k}_{aT}
+ \bbox{k}_{bT} -\bbox{q}_T)\nonumber\\
&&\times\text{Tr} \left[ \left(\int dk_a^-\Phi_a (k_a)\right) \gamma^\mu
\left(\int dk_b^+\Phi_b(k_b)\right) \gamma^\nu \right] ,\label{Horn}
\end{eqnarray}
rendering $x_a=x_A$ and $x_b=x_B$.
The leading order result
comes from inserting projections, Eqs.~(\ref{tweetwist}),
and the corresponding ones for hadron~$B$,
which can be obtained from the former by replacing $+\leftrightarrow -$,
$f^a\rightarrow f^{\bar{a}}$, $g^a\rightarrow -g^{\bar{a}}$, and
$h^a\rightarrow h^{\bar{a}}$~\cite{tang94}. One finds
\begin{eqnarray}
W_{\text{B}}^{\mu\nu} =-\frac{1}{3}\sum_{a,b}  \delta_{b\bar{a}}e^2_a &&
\int d^2\bbox{k}_{aT}\, d^2\bbox{k}_{bT}\; \delta^2(\bbox{k}_{aT}
+\bbox{k}_{bT}-\bbox{q}_{T})\nonumber\\
&&\times\biggl\{\biggl[ \Phi_{a}[\gamma^+]\,\Phi_{b}[\gamma^-]+
\Phi_{a}[\gamma^+\gamma_5]\,\Phi_{b}[\gamma^-\gamma_5]
\Bigr]g_T^{\mu\nu}
\nonumber\\&& \quad
+\Phi_{a}[i\sigma^{i+}\gamma_5]\,\Phi_{b}[i\sigma^{j-}
\gamma_5]\left(g_{T i}\!^{\,\{\mu}g_{T}\!^{\nu\}}\!_j
-g_{T ij}g_T^{\mu\nu}\right)\biggr\}+\ldots , \label{dots}
\end{eqnarray}
This expression still has the full dependence on $Q$, $x_A$, $x_B$, and $Q_T$.
If one does not measure $Q_T$, one has to take the $\bbox{q}_T$-integral of
the cross section~(\ref{cross}). In the dimuon rest frame ${\cal O}'$,
the lepton axis vector $\hat{l}$ has only spatial components, and does not
depend on $\bbox{q}_T$ anymore. Neither does $\hat{q}$, so the
$\bbox{q}_T$-integral may be
pulled through the lepton tensor to work on $W^{\mu\nu}$ directly.
We will call the result after integration $\overline{W}^{\mu\nu}$.
In the frame  ${\cal O}'$, the components of $P_A$ and $P_B$ (or alternatively
$n_+$ and $n_-$) have a $\bbox{q}_T$-dependence.
However, the vector $Z$ does not.

Returning to Eq.~(\ref{dots}), we may in leading order replace $g_T^{\mu\nu}$
by $g_\perp^{\mu\nu}$ according to Eq.~(\ref{gtgp}).
Integrating over $\bbox{q}_T$
in the dimuon rest frame ${\cal O}'$, the $g_\perp$'s may be pulled outside
the integral, since they are built from $\hat{q}$ and $\hat{z}$
[Eq.~(\ref{geperp})]. Now the only $\bbox{q}_T$-dependence resides in the
delta-function, which is subsequently cancelled. The transverse momentum
integrals then work on their corresponding projections directly, hence
we may use the parametrizations~(\ref{goedtwee}).
The result is the zeroth-order DY hadron tensor~\cite{rals79}
\begin{eqnarray}
\overline{W}^{\mu\nu}_{\text{B}0}=-\frac{1}{3}\sum_a e_a^2 &&
\Bigl\{\left[f_1^a(x_A)f_1^{\bar{a}}(x_B)-\lambda_A\lambda_B g_1^a(x_A)
g_1^{\bar{a}}(x_B)\right]g_\perp^{\mu\nu}\nonumber\\
&&+h_1^a(x_A)h_1^{\bar{a}}(x_B)
\left[S_{A\perp}^{\{\mu}S_{B\perp}^{\nu\}}-(S_{A\perp}\cdot S_{B\perp})
g_\perp^{\mu\nu}\right]\Bigr\},\label{DYB0}
\end{eqnarray}
where we used the symmetrization of indices,
$S_{A\perp}^{\{\mu}S_{B\perp}^{\nu\}}=S_{A\perp}^{\mu}S_{B\perp}^{\nu}
+S_{A\perp}^{\nu}S_{B\perp}^{\mu}$.

Since in this paper we are interested in subleading order,
we must find the ${\cal O}(1/Q)$ content of Eq.~(\ref{dots}).
The corrections that we neglected above by replacing  $g_T^{\mu\nu}$
by $g_\perp^{\mu\nu}$ are
\begin{eqnarray}
W_{\text{B1-1}}^{\mu\nu}
= -\frac{1}{3Q}\sum_{a,b}  \delta_{b\bar{a}}\,&&e^2_a
\int d^2\bbox{k}_{aT}\, d^2\bbox{k}_{bT}\; \delta^2(\bbox{k}_{aT}
+\bbox{k}_{bT}-\bbox{q}_{T})\nonumber\\
&&\times\biggl\{\biggl[ \Phi_{a}[\gamma^+]\,\Phi_{b}[\gamma^-]+
\Phi_{a}[\gamma^+\gamma_5]\,\Phi_{b}[\gamma^-\gamma_5]
\biggr]\hat{q}^{\{\mu}q_\perp^{\nu\}}
\nonumber\\&&
+\Phi_{a}[i\sigma^{i+}\gamma_5]\,\Phi_{b}[i\sigma^{j-}\gamma_5]
\hat{q}^{\{\mu}\left( g_{\perp i}\!^{\,\nu\}}q_{\perp j}
+g_{\perp j}\!^{\,\nu\}}q_{\perp i}-g_{T ij}q_\perp^{\nu\}}\right)\biggr\}.
\end{eqnarray}
Integrating over $\bbox{q}_T$ in the frame ${\cal O}'$, we may pull
$\hat{q}$ outside the integral. Aside from the delta function, the integrand
is then linear in $\bbox{q}_T$, selecting $\bbox{k}_T$-{\em odd\/} structures
in the projections, as can be seen from the resulting expression
\begin{eqnarray}
\overline{W}_{\text{B1-1}}^{\mu\nu} =
\frac{1}{3Q}\sum_a e_a^2\Biggl\{ &&\lambda_A
\,\hat{q}^{\{\mu}S_{B\perp}^{\nu\}} \;
g_1^a(x_A)\left[\int d^2\bbox{k}_T \frac{\bbox{k}_T^2}{2M_B}
g_{1T}^{\bar{a}}(x_B,\bbox{k}_T^2)\right]\nonumber\\
&&+\lambda_A
\,\hat{q}^{\{\mu}S_{B\perp}^{\nu\}} \;
\left[\int d^2\bbox{k}_T \frac{\bbox{k}_T^2}{M_A}
h_{1L}^{a\perp}(x_A,\bbox{k}_T^2)\right]h_1^{\bar{a}}(x_B)\nonumber\\
&&+\lambda_B
\,\hat{q}^{\{\mu}S_{A\perp}^{\nu\}} \;
\left[\int d^2\bbox{k}_T \frac{\bbox{k}_T^2}{2M_A}
g_{1T}^{a}(x_A,\bbox{k}_T^2)\right]g_1^{\bar{a}}(x_B)\nonumber\\
&&+\lambda_B
\,\hat{q}^{\{\mu}S_{A\perp}^{\nu\}} \;
h_1^{a}(x_A)\left[\int d^2\bbox{k}_T \frac{\bbox{k}_T^2}{M_B}
h_{1L}^{\bar{a}\perp}(x_B,\bbox{k}_T^2)\right]\Biggr\}.\label{B1tw2}
\end{eqnarray}
In addition, Eq.~(\ref{Horn}) contains contributions
which are convolutions of a leading and subleading
projection. The corresponding Lorentz tensors do not contain
explicit $\bbox{q}_T$-dependence in the photon rest frame, so
that the delta function vanishes
and the $\bbox{k}_{aT}$- and $\bbox{k}_{bT}$-integrals work on the
corresponding projections directly.
Using Eqs.~(\ref{goedtwee}) and~(\ref{knar}), we get
\begin{eqnarray}
\overline{W}_{\text{B1-2}}^{\mu\nu} =
\frac{1}{3Q}\sum_a e_a^2\Bigl\{ &&-M_B\lambda_A
(\hat{q}+\hat{z})^{\{\mu}S_{B\perp}^{\nu\}} \;
g_1^a(x_A)\,x_B g_T^{\bar{a}}(x_B)\nonumber\\
&&+M_A\lambda_A
(\hat{q}-\hat{z})^{\{\mu}S_{B\perp}^{\nu\}} \;
x_A h_L^a(x_A) h_1^{\bar{a}}(x_B)\nonumber\\
&&-M_A\lambda_B
(\hat{q}-\hat{z})^{\{\mu}S_{A\perp}^{\nu\}} \;
x_A g_T^a(x_A) g_1^{\bar{a}}(x_B)\nonumber\\
&&+M_B\lambda_B
(\hat{q}+\hat{z})^{\{\mu}S_{A\perp}^{\nu\}} \;
h_1^{a}(x_A)\, x_B h_L^{\bar{a}}(x_B)\Bigr\}.\label{dir}
\end{eqnarray}
If we eliminate $g_T(x)$ and $h_L(x)$ using
relations~(\ref{eqgt}) and~(\ref{eqhl}), respectively, like
in the preceding section, we obtain
$\overline{W}_{\text{B1}}^{\mu\nu}$ =
$\overline{W}_{\text{B1-1}}^{\mu\nu} +
\overline{W}_{\text{B1-2}}^{\mu\nu}$,
\begin{eqnarray}
\overline{W}_{\text{B}1}^{\mu\nu} =
\frac{1}{3Q}\sum_a e_a^2\biggl\{ &&-\lambda_A
\hat{z}^{\{\mu}S_{B\perp}^{\nu\}} \;
g_1^a(x_A)\left[\int d^2\bbox{k}_T \frac{\bbox{k}_T^2}{2M_B}
g_{1T}^{\bar{a}}(x_B,\bbox{k}_T^2)\right]\nonumber\\
&&+\lambda_A
\hat{z}^{\{\mu}S_{B\perp}^{\nu\}} \;
\left[\int d^2\bbox{k}_T \frac{\bbox{k}_T^2}{M_A}
h_{1L}^{a\perp}(x_A,\bbox{k}_T^2)\right]h_1^{\bar{a}}(x_B)\nonumber\\
&&+\lambda_B
\hat{z}^{\{\mu}S_{A\perp}^{\nu\}} \;
\left[\int d^2\bbox{k}_T \frac{\bbox{k}_T^2}{2M_A}
g_{1T}^{a}(x_A,\bbox{k}_T^2)\right]g_1^{\bar{a}}(x_B)\nonumber\\
&&-\lambda_B
\hat{z}^{\{\mu}S_{A\perp}^{\nu\}} \;
h_1^{a}(x_A)\left[\int d^2\bbox{k}_T \frac{\bbox{k}_T^2}{M_B}
h_{1L}^{\bar{a}\perp}(x_B,\bbox{k}_T^2)\right]\nonumber\\
&&+2m\left[\lambda_B \hat{z}^{\{\mu}S_{A\perp}^{\nu\}} \;
h_1^{a}(x_A) g_1^{\bar{a}}(x_B)
-\lambda_A\hat{z}^{\{\mu}S_{B\perp}^{\nu\}} \;
g_1^a(x_A) h_1^{\bar{a}}(x_B)\right]\nonumber\\
&&-M_B\lambda_A(\hat{q}+\hat{z})^{\{\mu}S_{B\perp}^{\nu\}} \;
g_1^a(x_A)\,x_B \tilde{g}_T^{\bar{a}}(x_B)\nonumber\\
&&+M_A\lambda_A
(\hat{q}-\hat{z})^{\{\mu}S_{B\perp}^{\nu\}} \;
x_A \tilde{h}_L^a(x_A) h_1^{\bar{a}}(x_B)\nonumber\\
&&-M_A\lambda_B
(\hat{q}-\hat{z})^{\{\mu}S_{A\perp}^{\nu\}} \;
x_A \tilde{g}_T^a(x_A) g_1^{\bar{a}}(x_B)\nonumber\\
&&+M_B\lambda_B
(\hat{q}+\hat{z})^{\{\mu}S_{A\perp}^{\nu\}} \;
h_1^{a}(x_A)\, x_B \tilde{h}_L^{\bar{a}}(x_B)\biggr\}.\label{DYB1}
\end{eqnarray}
As expected, the $\hat{q}$-terms that violate current conservation
only come multiplied with the explicitly interaction-dependent matrix elements
$\tilde{g}_T(x)$ and $\tilde{h}_L(x)$. In QCD, one must include the gluon
diagrams of Fig.~\ref{fig:DYgluon}. Consider first diagrams~(a)
and~(b) where the transverse gluon is inserted in hadron~$A$,
\begin{eqnarray}
W_{\text{gluon-}A}^{\mu\nu} =\frac{1}{3}\sum_{a,b}  \delta_{b\bar{a}}e^2_a&&
\int d^4k_a\, d^4k_b\, d^4 p\,\delta^4(k_a + k_b -q)\nonumber\\
&&\times\Biggl\{
\text{Tr} \left[ {\cal M}^i_{a} (k_a,p)\, \gamma^\mu\, \frac{i}{{\not\! p}
-{\not\! q}-m}\, i\gamma_i\,\Phi_{b}(k_b)\, \gamma^\nu \right]\nonumber\\
&&+ \text{Tr} \left[{\cal M}^i_{a} (p,k_a)\, \gamma^\mu\,\Phi_{b}(k_b)
 \, i\gamma_i\, \frac{i}{{\not\! p}-{\not\! q}-m} \,\gamma^\nu\right]\Biggr\}
\label{A-gluon}.
\end{eqnarray}
The delta function can be approximated as in Eq.~(\ref{delta}).
Since the leading projections $\Phi_b[\Gamma]$ are multipied by
an overall $\gamma^+$, and since
$\gamma^+\gamma^i\gamma^+ = 0$, the hard propagator effectively reduces to
\begin{equation}\label{DYprop}
\frac{1}{{\not\! p}-{\not\! q}-m}\rightarrow - \frac{{\not\! n_+}}{Q\sqrt{2}}.
\end{equation}
Therefore, the $p$-integral can be pulled through the trace to  work directly
on the quark-gluon-quark correlations. Using the hermiticity
condition~(\ref{Mcon}) to relate ${\cal M} (p,k_a)$ to ${\cal M} (k_a,p)$,
one is sensitive to the projections~(\ref{Mproje}) with
$\Gamma=\gamma^+$, $\gamma_5\gamma^+$,
or $i\sigma^{i+}\gamma_5$, or their hermitian conjugates.
Working out the details, one find that to ${\cal O}(1/Q)$ accuracy the
Lorentz structure of the expression~(\ref{A-gluon}) in the frame ${\cal O}'$
has no $\bbox{q}_T$-dependence, allowing us to use $\bbox{k}_T$-averaged
correlation functions in Eqs~(\ref{gTtwiddle}) and~(\ref{hLtwiddle}).
The result is
\begin{eqnarray}
\overline{W}_{\text{gluon-}A}^{\mu\nu} \approx
\frac{1}{3Q}\sum_a e_a^2\Bigl\{
&&-M_A\lambda_A
(\hat{q}+\hat{z})^{\{\mu}S_{B\perp}^{\nu\}} \;
x_A \tilde{h}_L^a(x_A) h_1^{\bar{a}}(x_B)\nonumber\\
&&+M_A\lambda_B
(\hat{q}+\hat{z})^{\{\mu}S_{A\perp}^{\nu\}} \;
x_A \tilde{g}_T^a(x_A) g_1^{\bar{a}}(x_B)\Bigr\}.
\end{eqnarray}
In the same way we find for the diagrams~\ref{fig:DYgluon}(c)
and~\ref{fig:DYgluon}(d) with a transverse gluon insertion in hadron~$B$:
\begin{eqnarray}
\overline{W}_{\text{gluon-}B}^{\mu\nu} \approx
\frac{1}{3Q}\sum_a e_a^2\Bigl\{
&&M_B\lambda_A(\hat{q}-\hat{z})^{\{\mu}S_{B\perp}^{\nu\}} \;
g_1^a(x_A)\,x_B \tilde{g}_T^{\bar{a}}(x_B)\nonumber\\
&&-M_B\lambda_B
(\hat{q}-\hat{z})^{\{\mu}S_{A\perp}^{\nu\}} \;
h_1^{a}(x_A)\, x_B \tilde{h}_L^{\bar{a}}(x_B)\Bigr\}.
\end{eqnarray}

The antiquark diagrams can be obtained from the corresponding
quark diagrams by replacing $m\rightarrow -m$, $e^a\rightarrow -e^{\bar{a}}$,
$f^a\rightarrow f^{\bar{a}}$, $g^a\rightarrow -g^{\bar{a}}$, and
$h^a\rightarrow h^{\bar{a}}$. However, the quark results are invariant under
this operation, so one can use them, extending the flavor sum to include
antiquarks.

The total ${\cal O}(1/Q)$ polarized DY hadron tensor contains no current
non-conserving terms anymore.
It can be written in terms of two structure functions,
\begin{equation}\label{W1}
\overline{W}_{\text{B}1}^{\mu\nu}
+\overline{W}_{\text{gluon-}A}^{\mu\nu}
+\overline{W}_{\text{gluon-}B}^{\mu\nu}=
\lambda_A\hat{z}^{\{\mu}S_{B\perp}^{\nu\}}\,\overline{U}^{LT}_{2,1}
-\lambda_B\hat{z}^{\{\mu}S_{A\perp}^{\nu\}}\,\overline{U}^{TL}_{2,1},
\end{equation}
with
\begin{eqnarray}
\overline{U}^{LT}_{2,1} =
-\frac{1}{3Q}\sum_a e_a^2\Biggl\{ &&
g_1^a(x_A)\left[\int d^2\bbox{k}_T \frac{\bbox{k}_T^2}{2M_B}
g_{1T}^{\bar{a}}(x_B,\bbox{k}_T^2)+2M_B x_B
\tilde{g}_T^{\bar{a}}(x_B)\right]\nonumber\\
&&+\left[-\int d^2\bbox{k}_T \frac{\bbox{k}_T^2}{M_A}
h_{1L}^{a\perp}(x_A,\bbox{k}_T^2)+2M_Ax_A
\tilde{h}_L^a(x_A)\right]h_1^{\bar{a}}(x_B)\nonumber\\
&&+2m\, g_1^a(x_A) h_1^{\bar{a}}(x_B)\Biggr\},\label{ULT}
\end{eqnarray}
where the sum runs over quark {\em and} antiquark flavors.
The other structure function, $\overline{U}^{TL}_{2,1}$, can be obtained from
Eq.~(\ref{ULT}) by replacing $A\leftrightarrow B$.
It is easily shown that they {\em cannot\/} be written in terms
of $g_T(x)$ and $h_L(x)$ only. However, using Eqs.~(\ref{eqgt})
and~(\ref{eqhl}), one {\em can\/} simultaneously eliminate the
explicit $\bbox{k}_T$-integrals and quark-mass term;
\begin{eqnarray}
\overline{U}^{LT}_{2,1} =
-\frac{1}{3Q}\sum_a e_a^2\Bigl\{ &&
M_B\,g_1^a(x_A)x_B\Bigl[g_T^{\bar{a}}(x_B)+\tilde{g}_T^{\bar{a}}(x_B)\Bigr]
\nonumber\\
&&+M_A\,x_A\Big[ h_L^a(x_A)+\tilde{h}_L^a(x_A)\Big]
h_1^{\bar{a}}(x_B)\Bigr\}.
\end{eqnarray}

\subsection{Discussion}

Having derived the DY hadron tensor, it is straightforward to calculate the
cross section and double-spin asymmetries after contracting the leptonic
tensor, written in terms of the angles $\theta$ and $\phi$,
with the hadronic tensor in Eq.~(\ref{W1}). The only nonzero ${\cal O}(1/Q)$
contributions come from $TL$ or $LT$ scattering, of which we
only consider the latter; the former simply follows by replacing
$A\leftrightarrow B$. The result is
\begin{equation}
\frac{d\sigma (\lambda_A,\bbox{S}_{BT})}{dx_A dx_B d\Omega}
=\frac{\alpha^2 }{4 Q^2}
\Bigl[(1+\cos^2\theta)\,\overline{W}_T+\lambda_A\sin 2\theta
\cos(\phi-\phi_B)\,
\overline{U}^{LT}_{2,1}\Bigr],\label{sigmalt}
\end{equation}
neglecting ${\cal O}(1/Q^2)$ contributions.
Here $\overline{W}_T= (1/3) \sum_a e_a^2\, f_1^a(x_A) f_1^{\bar{a}}(x_B)$ is
the leading-order unpolarized structure function. As transverse
momentum has been integrated over, this result depends only on the
relative azimuthal angle, $\phi-\phi_B$, between the
lepton scattering plane and the transverse polarization vector.
The asymmetry following from Eq.~(\ref{sigmalt}) reads
\begin{equation}
A_{LT}=\frac{\sigma(\lambda_A,\bbox{S}_{BT})-\sigma(\lambda_A,-\bbox{S}_{BT})}
{\sigma(\lambda_A,\bbox{S}_{BT})+\sigma(\lambda_A,-\bbox{S}_{BT})}=
\lambda_A\frac{\sin 2\theta \cos(\phi-\phi_B)}{1+\cos^2\theta}
\frac{\overline{U}^{LT}_{2,1}}{\overline{W}_T}.
\end{equation}
Taking $\lambda_A=-1$, $\phi_B=0$, and writing out the structure functions,
this becomes
\begin{eqnarray}
&&A_{LT}=
\frac{\sin 2\theta \cos\phi}{1+\cos^2\theta}\frac{1}{Q}\nonumber\\
&&\qquad\times\frac{\sum_a e_a^2\left\{
M_B\,g_1^a(x_A)x_B\Bigl[g_T^{\bar{a}}(x_B)+\tilde{g}_T^{\bar{a}}(x_B)\Bigr]
+M_A\,x_A\Bigl[ h_L^a(x_A)+\tilde{h}_L^a(x_A)\Bigr]
h_1^{\bar{a}}(x_B)\right\}}{\sum_a e_a^2\, f_1^a(x_A) f_1^{\bar{a}}(x_B)}.
\label{spruit}\end{eqnarray}
We emphasize again that Eq.~(\ref{spruit}) is the general result for the
${\cal O}(1/Q)$ LT-asymmetry. No assumptions have been made for the
$\bbox{k}_T$-dependence of the distributions.
If we assume $\delta(\bbox{k}_T^2)$ transverse momentum distributions and
$m=0$,
the relations~(\ref{eqgt}) and~(\ref{eqhl})
reduce to $g_T(x)=\tilde{g}_T(x)$ and $h_L(x)=\tilde{h}_L(x)$, respectively.
Inserting these into Eq.~(\ref{spruit}), we arrive at the
result derived by Jaffe and Ji in Ref.~\cite{jaff92} (taking $M_A=M_B=M$).

The bag model gives an estimate for the quark distributions.
Using the calculations of Refs.~\cite{jaff91a,jaff92} (see Sec.~\ref{sec2}),
we plot in Fig.~\ref{fig:3D} the ($x_A,x_B$)-dependent part (without
the angular dependence and the factor $M/Q$) of the
asymmetry~(\ref{spruit}). It turns out that the zero-transverse-momentum
zero-$m$ approximation is of roughly the same form, but shifted upwards.
This can also be seen from the cut along the line $x_A=x_B$, depicted in
Fig.~\ref{fig:proj}. Clearly, the two results differ considerably.
This could have been foreseen from the observation, made in Sec.~\ref{sec2},
that the  zero-transverse-momentum
zero-$m$ relations $g_T(x)=\tilde{g}_T(x)$ and $h_L(x)=\tilde{h}_L(x)$
in the bag are severely violated.

In sum, we have carefully analyzed the twist-three polarized distributions
$g_T(x)$ and $h_L(x)$ as to their dynamical content in a nonstandard (i.e.,
non-OPE) way. They are sensitive to nonzero intrinsic transverse
momentum, as is also clear from the analysis of the physical processes;
polarized deep-inelastic and Drell-Yan scattering at ${\cal O}(1/Q)$.
For the latter process we found that the LT-asymmetry, in addition
to $g_T(x)$ and $h_L(x)$, contains the
quark-gluon-quark correlations $\tilde{g}_T(x)$ and $\tilde{h}_L(x)$.

\acknowledgements

We would like to express our gratitude to the European Centre for Theoretical
Studies in Nuclear Physics and Related Areas (ECT*) in Trento, Italy for their
kind hospitality. We thank E. Leader for some useful suggestions.
This work was supported by the Foundation for Fundamental Research on Matter
(FOM) and the National Organization for Scientific Research (NWO).

\begin{figure}
\caption{\label{fig:quarkblob}The blob representing the quark-quark correlation
function $\Phi_{\alpha\beta}(PS;k)$.}
\end{figure}

\begin{figure}
\caption{\label{fig:gluonblob}The quark-gluon-quark correlation function
${\cal M}^i_{\alpha\beta}(PS;k,p)$.}
\end{figure}

\begin{figure}
\caption{\label{fig:physarea}The physical area in the $\sigma\tau$-plane
for $x=1/2$. The lower boundary is given by the line $\tau=\sigma-1$, the upper
by $\tau=x\sigma-x^2$.}
\end{figure}

\begin{figure}
\caption{\label{fig:gete}Distribution function $g_T(x)$ for a massless quark in
the bag (solid line), its twist-two Wandzura-Wilczek part
$\int_x^1 dy[g_1(y)/y]$ (dotted) and its interaction-dependent part
$\tilde{g}_T(x)-\int_x^1 dy [\tilde{g}_T(y)/y]$
(dashed)~\protect\cite{jaff91a}.
The dot-dashed line is $\tilde{g}_T(x)$ which diverges like $1/x$ for
$x\rightarrow 0$.}
\end{figure}

\begin{figure}
\caption{\label{fig:ghtwee}$h_2$ (solid curve) in the bag is twice $g_2$
(dashed).}
\end{figure}

\begin{figure}
\caption{\label{fig:hael} $h_L(x)$ for a massless quark in the bag (solid
line),
its twist-two part $2x\int_x^1 dy [h_1(y)/y^2]$ (dotted), and its
interaction-dependent part $\tilde{h}_L(x)-2x\int_x^1 dy [\tilde{h}_L(y)/y^2]$
(dashed)~\protect\cite{jaff92}. The distribution $\tilde{h}_L(x)$ (dot-dashed)
diverges like $1/x$ for $x\rightarrow 0$. }
\end{figure}

\begin{figure}
\caption{\label{fig:DISBorn}The quark Born diagram for DIS.}
\end{figure}

\begin{figure}
\caption{\label{fig:DISgluons}Gluon diagrams contributing to ${\cal O}(1/Q)$
DIS.}
\end{figure}

\begin{figure}
\caption{\label{fig:DYBorn}DY quark Born diagram.}
\end{figure}

\begin{figure}
\caption{\label{fig:DYgluon}DY quark diagrams with one transverse-gluon
insertion.}
\end{figure}

\begin{figure}
\caption{\label{fig:3D}$(x_A,x_B)$-dependent part of $A_{LT}$
[Eq.~(\protect\ref{spruit})] in the bag.}
\end{figure}

\begin{figure}
\caption{\label{fig:proj}The bag-model $(x_A,x_B)$-dependent part of $A_{LT}$
for $x_A=x_B$ (solid), and its ze\-ro-trans\-verse-mo\-men\-tum approximation
(dashed).}
\end{figure}

\end{document}